\documentclass[preprint,showpacs,aps]{revtex4}

\usepackage{graphicx}
\usepackage{dcolumn}
\usepackage{bm}

\begin{document}

\begin{titlepage}

\title{Overcoming Doping Difficulty in Graphene via Substrate: A First Principle Study}

\author{Bing Huang$^1$, H. J. Xiang$^2$, and Su-Huai Wei$^1$}

\address{$^1$National Renewable Energy Laboratory, 1617 Cole Boulevard, Golden, CO 80401, USA}

\address{$^2$Key Laboratory of Computational Physical Sciences and Department of Physics, Fudan University, Shanghai 200433, P. R. China}

\date{\today}

\begin{abstract}

Controlling the type and density of charge carriers by doping is the
key step for developing graphene electronics. However, direct doping
of graphene is rather challenge. Based on first-principles
calculations, a concept of overcoming doping difficulty in graphene
via substrate is reported. We find that doping could be strongly
enhanced in epitaxial graphene grown on silicon carbide substrate.
Compared to free-standing graphene, the formation energies of the
dopants can decrease by as much as 8 eV. The type and density of the
charge carriers of epitaxial graphene layer can be effectively
manipulated by suitable dopants and surface passivation. More
importantly, contrasting to the direct doping of graphene, the
charge carriers in epitaxial graphene layer are weakly scattered by
dopants due to the spatial separation between dopants and conducting
channel. Finally, we show that similar idea can also be used to
control magnetic properties, \emph{e.g.}, induces a half-metallic
state in the epitaxial graphene without magnetic impurity doping.

\end{abstract}


\maketitle

 \draft

\vspace{2mm}

\end{titlepage}

The conventional silicon based microelectronics is expected to
encounter fundamental limitations at nanoscale. According to the
semiconductor industry road map, novel materials that could
complement or substitute silicon are needed. It is shown that
graphene is suitable for coherent nanoscale electronics applications
due to its unique electronic properties and extremely high carrier
mobility\cite{Geim-2007, Castro Neto-2009, N. M. R. Peres-2010}.
Controlling the type and density of charge carriers by doping is at
the heart of graphene electronics. However, the development of
reliable chemical doping methods in graphene is still a real
challenge. Direct doping of graphene with substitutional
dopants is rather difficult, because the strong covalent C-C
$\sigma$ bonds make graphene as one of the strongest materials in
the world\cite{Geim-2007}. Until now, doping of graphene has mainly
been achieved through reactive molecular or atomic adsorbates in
experiments\cite{T. Ohta-2006, S. Y. Zhou-2008}, which is difficult
to implement under current device technology. Moreover,
direct doping usually suppresses the high carrier mobility of graphene
evidently\cite{Geim-2007, Castro Neto-2009, N. M. R. Peres-2010}.
Thus, new proposals for overcoming the doping difficulty in graphene
is urgently needed.

Comparing to mechanical exfoliation of graphene from graphite, large
area epitaxial growth of graphene layers on silicon carbide (SiC)
surface shows huge application potentials and has apparent
technological advantages over the mechanical exfoliation
method\cite{C. Berger-2006, M. Orlita-2008}. In this letter, we
demonstrate that substitutional doping could be strongly enhanced in
epitaxial graphene grown on SiC substrate comparing to that of
free-standing graphene (FSG). The formation energies of dopants can
reduce by as much as $\sim$ 8 eV. The type and density of carriers
in epitaxial graphene layer could be effectively manipulated by
suitable dopants and surface passivation. Contrasting to direct
doping of FSG, the carriers of epitaxial graphene are weakly
scattered by dopants due to the spatial separation between
scattering centers (dopants) and conducting channel (epitaxial
graphene). Thus, the intrinsic high carrier mobility of graphene
could be effectively maintained even during doping process. Finally,
we show that the same idea also can be used to control spin
properties of epitaxial graphene by buffer layer defects. We find
that the reconstructed defects in buffer layer can break the spin
symmetry of epitaxial graphene and half-metallicity can be induced
in the system without magnetic impurity doping.

All the calculations are performed using the method based on density
functional theory in the local spin density approximation (LSDA) as
implemented in the VASP code\cite{VASP}, which is known to describe
the structural and electronic properties of graphite quite
well\cite{A. Mattausch-2007, S. Kim-2008}. The LSDA results are
further checked by the generalized gradient approximation and the
results are consistent with each other. The electron-ion interaction
is described by the projector augmented wave method, and the energy
cutoff is set to 400 eV. Structural optimization is carried out on
all systems until the residual forces are converged to 0.02 eV/\AA.
A $\sqrt 3 \times \sqrt 3 R30^o$ SiC substrate cell is used to
accommodate a 2 $\times$ 2 epitaxial graphene cell, approximating to
the larger reconstruction with a $6\sqrt 3 \times 6\sqrt 3 R30^o$
periodicity usually observed in experiments. This approximation has
been proved to well describe the electronic properties of epitaxial
graphene on SiC\cite{F. Varchon-2007, A. Mattausch-2007}. More
importantly, the $\sqrt 3 \times \sqrt 3 R30^o$ reconstruction is
also been observed in experiments during epitaxial growth of
graphene on both (C-terminated) SiC(000$\overline 1$)
surface\cite{C. Berger-2006, J. Hass-2006} and (Si-terminated)
SiC(0001) surface\cite{Forbeaux-1998, Rutter-2007, Tromp-2009}. To
address the doping effects in graphene-SiC system, a $2\sqrt 3
\times 2\sqrt 3 R30^o$ SiC substrate cell, which accommodates a 4
$\times$ 4 graphene cell, is constructed. The SiC substrate is
modeled by a slab contains 4 SiC bilayers with H passivation on the
second surface of the slab. Several test calculations on 6 bilayers
and 8 bilayers substates give essentially the same results. A gamma
centered 6 $\times$ 6 $\times$ 1 \textbf{k}-point sampling is used
for the Brillouin-zone integration, including the Dirac point. In
the following, although our discussion is focused on the results of
graphene on SiC(000$\overline 1$) surface, the results for graphene
on SiC(0001) surface will also be mentioned.

\begin{figure}[tbp]
\includegraphics[width=8.0cm]{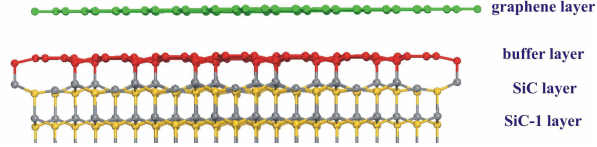}
\caption{(color online) The side view of optimized structure of
graphene layers on SiC(000$\overline 1$) surface. The top epitaxial
graphene layer and the rippled interface buffer layer are
highlighted in green/light gray and red/dark gray, respectively.
Only half of the SiC substrate is plotted here.}
\end{figure}

The optimized structure of graphene layers on SiC(000$\overline 1$)
is shown in Fig. 1. 2/3 of the C atoms in the top SiC surface make
covalent bonds with the C atoms of the interface buffer layer (the
first graphene layer during vacuum graphitization on SiC), with a
C-C bond length $\sim$ 1.62 \AA, which drastically changes the
Dirac-like band structures of the buffer layer, agreeing with
previous results\cite{F. Varchon-2007, A. Mattausch-2007, S.
Kim-2008}. A similar buffer layer structure is also observed on
SiC(0001) surface. The second graphene layer on the top of buffer
layer is planar and the distance between the rippled buffer layer
and top graphene layer varies between 3.34 - 3.71 \AA. The interface
buffer layer is stretched by $\sim$ 8\% as a result of the lattice
mismatch between $2\sqrt 3 \times 2\sqrt 3 R30^o$ SiC surface and
graphene, agreeing with previous predictions\cite{F. Varchon-2007,
A. Mattausch-2007, S. Kim-2008}.

It is expected that the carriers in graphene could be manipulated by
p/n-type dopants, like in conventional semiconductors\cite{A.
Franceschetti-1999}. Al, B, P, and N are considered as
substitutional dopants, as they are widely used in C- and Si- based
materials. All the possible substitutional positions in graphene-SiC
system are checked and the formation energies of dopants at
different positions are calculated to search the most stable
configurations. The formation energy ($E_f$) of a substitutional
dopant in graphene-SiC system is defined as:
\begin{equation}
E_f  = E_{doped}  - E_{host}  + \mu _{C/Si}  - \mu _{dopant}
\end{equation}
where $E_{doped}$ and $E_{host}$ are the total energies of doped and
undoped system, respectively. $\mu _{C/Si}$ and $\mu _{dopant}$
(dopant = Al, B, P, N) are the chemical potentials of C (or Si) and
dopant, respectively. Since epitaxial graphene should grow in C-rich
environment, $\mu _{C}$ is calculated assuming epitaxial graphene is
stable and $\mu_{Si}$ is calculated so that SiC is stable. It should
be noticed that the relative stability between different doping
positions does not depend on the particular choice of $\mu
_{dopant}$. The calculated formation energy differences are shown in
Fig. 2. Interestingly, Al prefers to substitute Si atom of the SiC
substrate or substitute C atom of the interface buffer layer and the
$E_f$ of Al decreases dramatically by $\sim$ 8.26 eV compared to FSG
case. B and P atoms prefer to substitute C atoms of the buffer layer
and the $E_f$ of dopants decrease largely by 2.36 eV and 5.27 eV,
respectively, compared to FSG cases. Besides doping in buffer layer,
the $E_f$ of B and P at SiC surface are also reduced by 1.46 eV and
4.87 eV, respectively, compared to FSG cases. Contrary to the cases
of Al, B, and P doping, N atom prefers to substitute C atom of top
SiC substrate with a reduction of $E_f$ by 3.52 eV, as shown in Fig.
2d. The similar phenomenon of substrate-enhanced doping is also
found on SiC (0001) surface, with the $E_f$ of dopants decreased by
as much as 10.8 eV.

\begin{figure}[tbp]
\includegraphics[width=8.0cm]{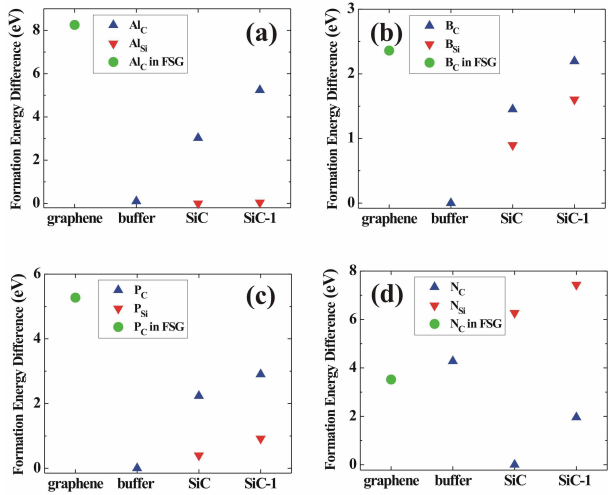}
\caption{(color online) The calculated formation energy differences
of dopants (a) Al, (b) B, (c) P, and (d) N in graphene-SiC system as
a function of different layers. The formation energies of dopants in
free-standing graphene (FSG) are also plotted in these figures for
comparison.}
\end{figure}

These quite different doping behaviors between different dopants as
well as different layers in graphene-SiC system could be understood
by considering the structural strain in the various epitaxial
graphene-SiC layers, the different atomic covalent radiuses, and the
electronegativity of dopants. Since the interface buffer layer is
under large tensible strain, Al, B, and P atoms could be easier to
substitute C atoms in this layer due to the larger covalent radiuses
of Al (1.21~\AA), B (0.84~\AA), P (1.07~\AA) than \emph{sp$^{2}$} C
(0.73~\AA). The rippled (deformed) structure reduces the doping
difficulty in buffer layer to some extent. Furthermore, because the
electronegativity of C (2.55) is larger than Si (1.90) in SiC, Si
behaves as positively charged cation and C acts as negatively
charged anion. Al atom has lower electronegativity (1.61) and a
little larger covalent radiuses than Si (1.11~\AA), so it would
prefer to substitute Si atom in the SiC matrix to increase the
negative Coulomb interaction between Al and C. Differing from Al, B,
and P, the structural tensile strain increases the difficulty of N
doping in the graphene layer because N has a smaller covalent radius
(0.71~\AA) than C. N atom has even higher electronegativity (3.04)
than C, so it would prefer to substitute C atom in the SiC matrix to
increase the negative Coulomb interaction between Si and N, which is
in agreement with the calculated trend of $E_f$ in Fig. 2d. More
importantly, the large reduction of $E_f$ for all the dopants at
strain-free SiC substrate (1.46 $\sim$ 8.26 eV) indicates that, not
only in the $2\sqrt 3 \times 2\sqrt 3 R30^o$ structure, the strongly
substrate-enhanced doping can also exist in $6\sqrt 3 \times 6\sqrt
3 R30^o$ reconstruction because the strain-free substrate is independent of reconstruction in the graphene or buffer layers. The above results and analysis show that
overcoming the doping difficulty in graphene could be generally
achieved through substrates, and it can also be applied to other
layered or thin-film systems. Moreover, it is quite encouraging to
see that highly strained graphene (up to 25\%) have already been
created on substrates in experiments through various schemes\cite{K.
S. Kim-2009, Levy-2010}.

After knowing the stable doping configurations, we turn to
investigate how these dopants tune the carrier properties,
\emph{i.e.} carrier type and carrier density,  in graphene-SiC
system. Without doping, the Dirac-like states in the top epitaxial
graphene layer on SiC (000$\overline 1$) is preserved, as shown in
Fig. 3a.  The weakly dispersive interface states visible in Fig. 3a
result from the surface dangling bonds of SiC substrate. In our unit
cell, there are twelve C \emph{sp$^{3}$} dangling orbitals on the
top layer of the SiC substrate. Eight of them make covalent bonds
with the buffer layer but the other four remain unsaturated. The
localization of C \emph{sp$^{3}$} dangling bonds favor the
spin-polarization and thus split to four occupied states and empty
states, respectively. Because the band alignment between the
epitaxial graphene states and substrate surface states is type-I,
therefore, there is no charge transfer between the epitaxial
graphene and substrate surface. However, the following results show
that the substrate surface states play an important role in
affecting the carrier density of top graphene layer.

\begin{figure}[tbp]
\includegraphics[width=6.0cm]{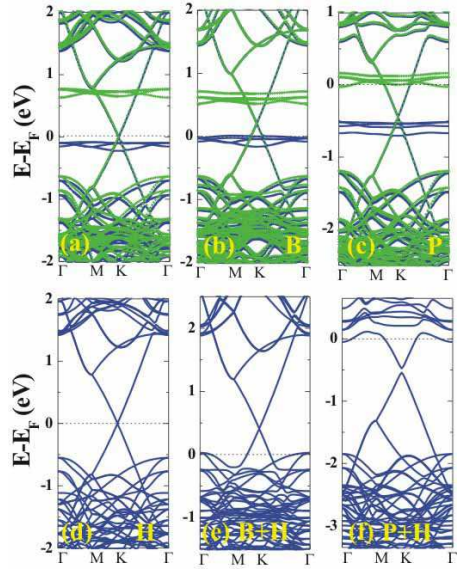}
\caption{(color online) The electronic band structures of
graphene-SiC system without doping (a) and doped by (b) B and (c) P
in buffer layer. (d)-(f) are the same as (a)-(c) but the C atoms
with dangling bonds at the SiC surface are passivated by H atoms.
The blue/dark gray solid line and green/light gray dotted line
represent the spin-up and spin-down states, respectively. The Fermi
level is set at zero energy.}
\end{figure}

Our total energy calculations indicate that dopants favor to locate
at buffer layer or close to buffer layer in the substrate. To
understand the carrier distribution, Bader analysis\cite{Bader} is
used to estimate the charge variation in the dopant and various
layers. We find that after substitutional doping in buffer layer, B
atom shows an onsite charge of 1.32 \emph{e}, whereas in the case of
Al doping, the value is substantially lower, 0.80 \emph{e}. The
variation between the Al and B cases could be understood by noticing
that B is much more electronegative than Al. When the total hole
carrier density is $\sim$ 1.02 $\times$ 10$^{14}$ cm$^{-2}$, the
hole density in the buffer layer is $\sim$ 5.77 $\times$ 10$^{13}$
cm$^{-2}$ after B doping and $\sim$ 9.98 $\times$ 10$^{13}$
cm$^{-2}$ after Al doping. Unexpectedly, the top epitaxial graphene
is not effectively doped by holes in either cases, with hole density
at $\sim$ 5 $\times$ 10$^{12}$ cm$^{-2}$. The origin of the small
carrier density in the epitaxial graphene layer can be understood
from the band structure of the graphene-SiC system. We see that
there are localized interface states which strongly pin the Fermi
level at about 0.18 eV below the Dirac point for $\emph{p}$-type
materials, as shown in Fig. 3b. (Al-doping case is similar and not
shown here). The hole density of epitaxial graphene is thus limited
to $\sim$ 5 $\times$ 10$^{12}$ cm$^{-2}$ for $\emph{p}$-type doped
system. For the $\emph{n}$-doped case (Fig. 3c), the empty interface
states also play a similar role and pin the Fermi level at about
0.44 eV above the Dirac point, which is responsible to the limited
electron density at $\sim$ 1.99 $\times$ 10$^{13}$ cm$^{-2}$ in the
top epitaxial graphene layer after P doping. The above results
strongly demonstrate that although B (Al) and P doping in buffer
layer could induce electrons or holes in the top epitaxial graphene,
the interface states play as a "subthreshold valve" to restrict the
carrier density. Similar doping behaviors are found when B (Al) and
P dope at SiC surface. As we mentioned, N atom prefers to substitute
the C atom with dangling bond at the top SiC layer. Contrary to P
doping, the epitaxial graphene layer could not be doped with
electrons because the N atom is strongly electronegative so the
spin-split states of N atom are both occupied after doping.

An obvious route to improve the electron or hole density in
epitaxial graphene is by eliminating the influence of the interface
states. Since these interface states originate from the dangling
orbitals of C atoms at the top of SiC substrate, naturally, we may
eliminate them by surface passivation. Here, we take H as an example
to demonstrate the surface passivation effect on modulating the
charge density of epitaxial graphene. In fact, H intercalation in
epitaxial graphene has already been achieved in several
experiments\cite{Riedl-2009, Guisinger-2009}. The main effect of H
adsorption is the disappearance of the interface states, as shown in
Fig. 3d. Without these interface states, we expect more electrons or
holes will transfer from buffer layer to epitaxial graphene layer.
The band structure calculations confirm our idea, as shown in Fig.
3e and 3f for B and P doping, respectively. After surface H
passivation, the hole density of B-doped system increases four times
up to 2.48 $\times$ 10$^{13}$ cm$^{-2}$, which pushes the the Fermi
level to 0.42 eV below the Dirac point. Similarly, the electron
density of P-doping system increases twice to 3.58 $\times$
10$^{13}$ cm$^{-2}$. Moreover, a small band gap ($\sim$ 0.07 eV)
appears in P-doping system, as shown in Fig. 3f. These results
strongly demonstrate that not only we can control the type of
carriers in epitaxial graphene via doping of the buffer layer, but
also the carrier density could be controlled by substrate
passivation. More noticeably, the dopants prefer to stay in the
buffer layer while the carriers are concentrated in the top
epitaxial graphene. Thus, the carriers in epitaxial graphene layer
could be only weakly scattered by dopants due to the spatial
separation between scattering centers (dopants) and conducting
channel (top graphene layer). This is in the same spirit as in
modulation doping in heterostructures such as GaAlAs/GaAs\cite{S.
Wang-1989}, and we suggest that it is an efficient way in
maintaining the high carrier mobility of graphene during doping
process.

\begin{figure}[tbp]
\includegraphics[width=8.0cm]{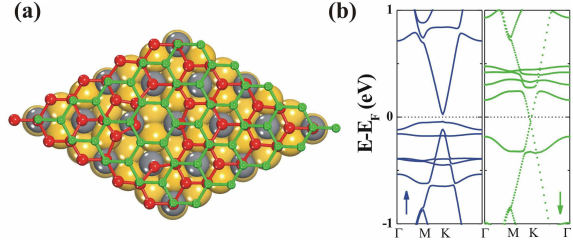}
\caption{(color online) (a) The top view of optimized structure of
graphene-SiC system with reconstructed vacancy in the buffer layer
(small red/dark gray atoms). (b) The electronic band structure of
(a). The blue/dark gray solid line and green/light gray dotted line
represent the spin-up and spin-down states, respectively. The Fermi
level is set at zero energy.}
\end{figure}

The manipulation of buffer layer not only can be used to control the
charge carriers, but also could be used to tune the spins of the
system. Here we demonstrate that the reconstructed vacancy in buffer
layer has unexpected effect on the spin properties of epitaxial
graphene. Injection of high spin-polarized current in graphene is
the current subject of intense investigation
efforts\cite{Tombros-2007}. Ideally, 100\% spin-polarized current
could be induced by half-metallic materials. Although spontaneous
vacancy in FSG may induce local spin-polarization, it is rare due to
its high formation energy of $\sim$ 8 eV. In graphene-SiC system, we
find that vacancy prefers to be formed in buffer layer and undergoes
reconstruction, as shown in Fig. 4a. A five-thirteen ring is formed
with a $E_f$ of 3.2 eV, which is $\sim$ 5 eV lower than in FSG. The
deformed buffer layer as well as the defect reconstruction are
responsible for the lower $E_f$ of vacancy in buffer layer. In
experiments, considerable amounts of defects have been observed in
the interface layer of epitaxial graphene system\cite{Rutter-2007,
Qi-2010}, especially the existence of reconstructed
hexagon-pentagon-heptagon\cite{Qi-2010}. The experimental
observations indicate that the $E_f$ of defect in buffer layer is
much lower than in FSG, agreeing with our results. It is interesting
to see that the reconstructed vacancy not only breaks the honeycomb
symmetry, but also breaks the spin symmetry of graphene, which
induces a half-metallic state in the top epitaxial graphene layer,
as shown in Fig. 4b. There is an apparent gap ($\sim$ 0.142 eV) in
the spin-up state of the epitaxial graphene and a negligible band
gap in spin-down state. The localized states near the band edge in
spin-up band structure are induced by the C atoms which are bonded
to the SiC surface in the five-thirteen ring. It was theoretically
predicted that half-metallicity could be induced in graphene by an
external electric field\cite{Son-2006} or transition metal
doping\cite{Jayasekera-2010}. However, the required electric field
is too strong and heavy transition metal elements also often act as
poison agents in biological systems. Our results here strongly
indicate that SiC-graphene system with reconstructed defects in the
buffer layer could have strong potentials for spintronics.

In conclusion, a concept of overcoming doping difficulty in graphene
via substrate is reported. We show that substitutional doping could
be strongly enhanced in epitaxial graphene on SiC substrate.
Compared to free-standing graphene, the formation energies of
dopants decrease by as much as 8 eV. The type and density of
carriers in epitaxial graphene could be effectively manipulated by
suitable dopants and surface passivation. The carriers in epitaxial
graphene layer are weakly scattered by dopants due to the spatial
separation between the dopants and carriers. Finally, we show that
the reconstructed vacancy in buffer layer could induce
half-metallicity in epitaxial graphene without magnetic impurity
doping. Generally, the effect of substrate-enhanced doping could
exist in other substrates, and it can also be applied to other
layered or thin-film systems.

The work at NREL was supported by the U.S. Department of Energy
under Contract No. DE-AC36-08GO28308. The work at Fudan was
partially supported by the National Science Foundation of China.

\newpage

\end{document}